\documentclass[article]{aa}
\usepackage[varg]{txfonts}

\usepackage{ulem}

\usepackage{hyperref}
\hypersetup{colorlinks=true,linkcolor=blue,citecolor=blue,filecolor=blue,urlcolor=blue}

\usepackage{tikz,xcolor,graphicx}
\graphicspath{{images/}}

\usepackage{bm}

\newcommand{\Msun}{M_\odot}
\newcommand{\Myr}{\mathrm{Myr}}
\newcommand{\kms}{\mathrm{km\,s^{-1}}}

\newcommand{\pc}{\mathrm{pc}}
\newcommand{\kpc}{\mathrm{kpc}}

\newcommand{\rh}[1][]{r_\mathrm{h#1}}
\newcommand{\rg}[1][]{x_\mathrm{#1}}

\newcommand{\Msgra}{M_\mathrm{SgrA*}}
\newcommand{\Mimbh}{M_\mathrm{IMBH}}
\newcommand{\SgrA}{Sgr\,A*}
\newcommand{\adeg}{^\circ}

\definecolor{lime}{HTML}{A6CE39}
\DeclareRobustCommand{\orcidicon}{%
	\begin{tikzpicture}
	\draw[lime, fill=lime] (0,0)
	circle [radius=0.16]
	node[white] {{\fontfamily{qag}\selectfont \tiny ID}};
	\draw[white, fill=white] (-0.0625,0.095)
	circle [radius=0.007];
	\end{tikzpicture}
	\hspace{-2mm}
}

\newcommand{\orcidVP}{\href{https://orcid.org/0000-0002-3031-062X}{\orcidicon}}
\newcommand{\orcidVK}{\href{https://orcid.org/0000-0002-5760-0459}{\orcidicon}}
\newcommand{\orcidBB}{\href{https://orcid.org/0000-0002-3578-6037}{\orcidicon}}
\newcommand{\orcidFP}{\href{https://orcid.org/0000-0002-9850-2708}{\orcidicon}}
\newcommand{\orcidAE}{\href{https://orcid.org/0000-0001-6049-3132}{\orcidicon}}

\begin{document}

\title{Dynamics of star associations in an SMBH--IMBH system}
\subtitle{The case of IRS13 in the Galactic centre}
\titlerunning{Dynamics of IRS13-like clusters}

\author{V\'aclav Pavl\'ik \inst{\ref{asu},\ref{iu},}\thanks{\email{pavlik@asu.cas.cz}} \orcidVP
\and Vladim\'ir Karas \inst{\ref{asu}} \orcidVK
\and Bhavana Bhat \orcidBB
\and Florian Pei{\ss}ker \inst{\ref{koln}} \orcidFP
\and Andreas Eckart \inst{\ref{koln},\ref{mpi}} \orcidAE
}

\institute{
	Astronomical Institute of the Czech Academy of Sciences, Bo\v{c}n\'i~II~1401, 141~00~Prague~4, Czech Republic \label{asu}
	\and Department of Astronomy, Indiana University, Swain Hall West, 727 E 3$^\text{rd}$ Street, Bloomington, IN 47405, USA \label{iu}
    \and
	I.~Physikalisches Institut der Universit\"at zu K\"oln, Z\"ulpicher Str.~77, D-50937 K\"oln, Germany \label{koln}
	\and
	Max Planck Institut f\"ur Radioastronomie, Auf dem H\"ugel 69, D-53121 Bonn, Germany \label{mpi}
}

\authorrunning{Pavl\'ik et al.}

\date{Received: August 30, 2024 / Accepted: October 18, 2024}

\abstract
{The existence of intermediate-mass black holes (IMBHs) still poses challenges to theoretical and observational astronomers. Several candidates have been proposed, including the one in the IRS13 cluster in the Galactic centre, where the evidence is based on the velocity dispersion of its members, however, none have been confirmed to date.}
{We aim to gain insights into the presence of an IMBH in the Galactic centre by a numerical study of the dynamical interplay between an IMBH and star clusters (SCs) in the vicinity of a supermassive black hole (SMBH).}
{We use high-precision $N$-body models of IRS13-like SCs in the Galactic centre, and of more massive SCs that fall into the centre of the Galaxy from larger distances.}
{We find that at IRS13's physical distance of $0.4\,\pc$, an IRS13-size SC cannot remain gravitationally bound even if it contains an IMBH of thousands $\Msun$. Thus, IRS13 appears to be an incidental present-day clumping of stars. Furthermore, we show that the velocity dispersion of tidally disrupted SCs (the likely origin of IRS13) can be fully accounted for by the tidal forces of the central SMBH; the IMBH's influence is not essential.}
{}


\maketitle

\section{Introduction}
\label{sec:intro}

The formation and growth of intermediate-mass black holes (IMBHs) have been studied intensively over the recent decade \citep[recently, e.g.][and citations there in]{ziosi, mocca_IMBH1, mocca_IMBH2, aros_binaries}. However, the existence of IMBHs still poses a challenge to astrophysical observations with no candidate being decisively confirmed thus far.
Besides massive globular clusters and the cores of dwarf galaxies, the presence of IMBHs has also been hypothesised in the Galactic centre; specifically inside IRS13 \citep{maillard_etal2004} which is reportedly a remnant core of a tidally disrupted star cluster (SC).
Using the proper motions of stars inside the IRS13 region, several authors have constrained the mass of this possible IMBH (originating within the infalling SC) to some thousands of Solar masses \citep{maillard_etal2004, schodel_etal2005, fritz_etal2010} to explain the relatively high stellar velocity dispersions. This estimate has recently been updated by \citet{peissker_etal2023} to $(3.9 \pm 0.1)\times 10^4\,\Msun$ based on the 2D stellar velocity dispersion of $(128.86 \pm 0.14)\,\kms$.

Numerical models by \citet{fujii_trojans} show that an IMBH (mass from $3{\times}10^3\,\Msun$ to $1.6{\times}10^4\,\Msun$) may form close to the central \SgrA\ supermassive black hole (SMBH) via collisions of stars that were initially in a cluster surrounding the SMBH. Furthermore, such an IMBH may help bring other massive stars to the vicinity of the central SMBH and form disks similar to the observed ones near \SgrA. However, using a more accurate stellar evolution prescription, \citet{petts_gualandris17} found that an IMBH would likely not grow past a few hundred $\Msun$. Consequently, it would not be massive enough to experience significant dynamical friction to drag stars close to \SgrA.

The presence of an IMBH in this region would also explain the X-ray emission which coincides with the source IRS13~E3 and the \textit{ALMA} observations which reveal an ionised ring of gas at that position \citep{tsuboi_etal2017,peissker_etal2024} --- assuming that the gas is virialized, the mass yields $3{\times}10^4\,\Msun$. A massive body should also ensure the dynamical stability of its surrounding stellar association since IRS13, containing ${\sim}50$ confirmed sources \citep{peissker_etal2023}, is only ${\approx}0.13\,\pc$ (projected) or ${\approx}0.4\,\pc$ (3D distance) from \SgrA\ \citep[see][]{tsuboi_etal2020}. Although the studies focusing solely on this cluster's dynamical development and lifetime are sparse, broader implications originating from the tidal dissolution of SCs near \SgrA\ were discussed \citep[see, e.g.][]{fujii_etal2008, as_gual2018, wang_lin2023}. Some SCs modelled by these authors end up in the Galactic centre reasonably fast (within a few Myr), with masses comparable to the physical properties of IRS13 \citep[see also][who estimate a shorter infall time]{Bonnell_Rice}.

\begin{figure}
    \centering
    \includegraphics[width=\linewidth]{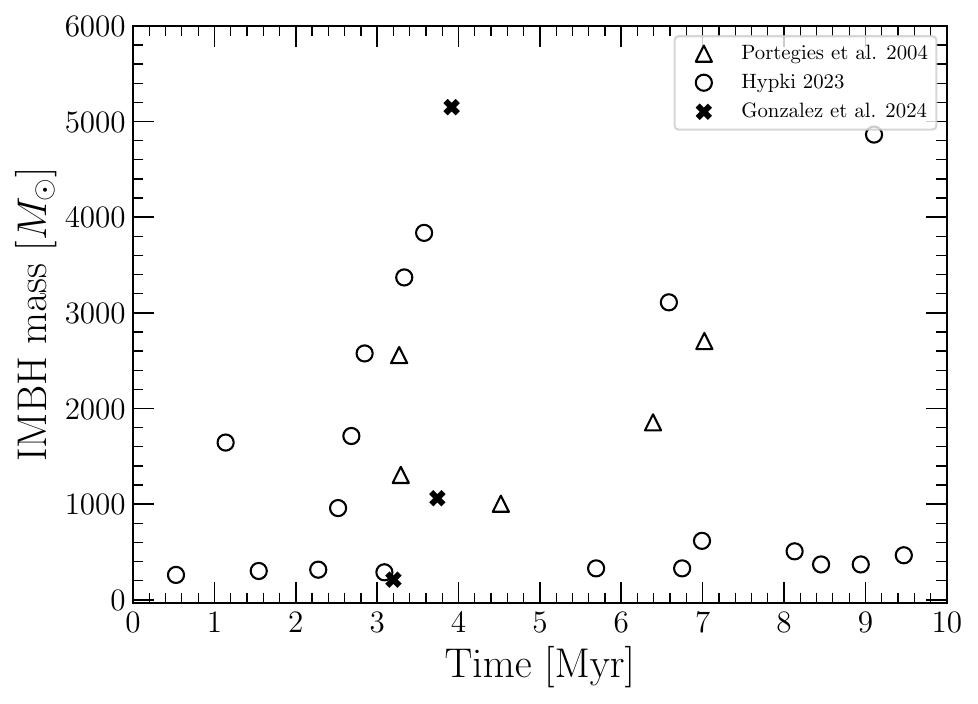}
    \caption{Time scales for the growth of IMBH masses by runaway mergers in star cluster models described in \cite{Portegies_zwart+2004}, \citet{hypki_ESO} and \citet{Gonzalez_prieto+2024} are shown with triangle, circle, and cross symbols, respectively.}
    \label{fig:imbh_literature}
\end{figure}

On the other hand, forming an IMBH of tens of thousands of Solar-masses takes typically an order of magnitude longer than the reported age of the IRS13 members \cite[${\sim}4\,\Myr$ for the first generation, see][]{peissker_etal2023}.
Among the several possible pathways, the two main scenarios for forming IMBHs have been envisaged: runaway mergers of main sequence stars, or accretion of residual gas onto stellar-mass black holes originating from first-generation stars.
Even in the best-case scenario based on the formation of very massive stars, runaway mergers produce IMBHs of only ${\sim}10^3\,\Msun$ within the first $10\,\Myr$ \citep{Portegies_zwart+2004, Gonzalez_prieto+2024, hypki_ESO}, see also Fig.~\ref{fig:imbh_literature}.%
\footnote{We note that several factors may influence the time scales plotted in Fig.~\ref{fig:imbh_literature}. First, the number of models analysed by \citet{Portegies_zwart+2004} is limited since the work focused on two particular young SCs (MGG-9 and MGG-11), which may not represent the variety of possible conditions for IMBH formation. Second, although \citet{Gonzalez_prieto+2024} explored many SC models with an IMBH, they only report its mass growth for some. In contrast, \citet{hypki_ESO} explored a wider set of models and reported the formation times of all IMBHs.}
Stellar densities required to build up an IMBH of mass ${\gtrsim}10^4\,\Msun$ are very high (${>}10^8\,\Msun\,\pc^{-3}$), as shown in the `FAST' scenario described by \cite{giersz_etal2015}, and are more common in nuclear clusters of small galaxies than in SCs (including globular clusters). In the gas accretion scenario, the IMBH mass has been shown to grow up to $10^4\,\Msun$, however, even with an accretion rate close to the Eddington rate the time scale is of the order of $20{-}40\,\Myr$ \citep{Vesperini+2010}.
 
Furthermore, \citet{gualandris_merritt2009} analysed the constraints on the mass and semi-major axis of a potential IMBH orbiting \SgrA\ with $N$-body models (based on the stability of the inner S-cluster) and an extensive review of the literature. They revealed that the existence of a massive object with $\Mimbh \gtrsim 2 \times 10^4\,\Msun$ at a distance of $0.4\,\pc$ from \SgrA\ is inconsistent with observations \citep[see also][]{rb2004}. The recent research by \citet{longwang_nature} also points out that an IMBH companion of \SgrA\ with the mass of $10^4\,\Msun$ is enough to explain the stellar kinematics in the inner parsec.

These works raise three main questions ---
(1) Does the hypothetical IMBH need to follow the same evolutionary path as IRS13 or can its position in this stellar association be coincidental?
(2) What role does an IMBH play against the continuous gravitational tides of an SMBH?
(3) Do we need an IMBH to explain the stellar motions within a tidally dissolved cluster?
--- and we address them in this paper by investigating the dynamical interplay between tidally disrupted stellar systems and an IMBH orbiting close to an SMBH.

\section{Orbiting clusters}
\label{sec:orbits}

\subsection{Models}

We first analysed a set of models of IRS13-like SCs (with 44 stars) to assess their stability while orbiting in the extreme tidal field of the central SMBH \citep[$\Msgra = 4.297 \times 10^6\,\Msun$; see][]{scluster_sgra}.
The stellar positions were randomly drawn from the \citet{plummer} distribution with exponential cut-off, so that the SC would have a maximal radius of ${\approx}0.07\,\pc$ \citep[i.e.\ the approximate size of IRS13 at the assumed Galactocentric distance of $8.5\,\kpc$; see][]{peissker_etal2023}.%
\footnote{We also tested other models where the initial distribution function was either uniform or \cite{king_model} with $W_0 = 3, 6$ and 9. However, due to the small number of stars, the results are comparable with the Plummer models we show.}
For simplicity, the SCs were generated with equal-mass stars (1 or $10\,\Msun$) and isotropic stellar velocity distribution.
The SCs were initialised with an IMBH in the centre; we first considered three values of its mass: $10^3\,\Msun$ \citep{maillard_etal2004}, $10^4\,\Msun$ \citep{schodel_etal2005}, and $4{\times}10^4\,\Msun$ \citep{peissker_etal2023, peissker_etal2024} to represent the proposed values for IRS13; then we also extended our analysis to a wider range of IMBHs.
We also varied the initial trajectory of the SC around the SMBH --- either circular with the radius of $0.4\,\pc$ or elliptical with the pericentre at this same distance and $0.5$ eccentricity. To compare, we also ran a set of models with isolated clusters, without an SMBH.

Numerically, the initial conditions were created using the \textsc{Agama} package \citep{agama} ensuring that the whole SC is bound and in virial equilibrium. Because of the large range of masses, we chose to integrate the models with the symplectic Wisdom--Holman integrator \textsc{WHFast} \citep{wh,reboundwhfast}, which is part of the \textsc{Rebound} $N$-body code \citep{rebound}. All objects (i.e.\ stars, the IMBH and the SMBH) were treated as point masses; general-relativistic effects, gas and dust were not implemented in these simulations but are planned for our follow-up research.

\begin{figure}
    \centering
    \includegraphics[width=\linewidth]{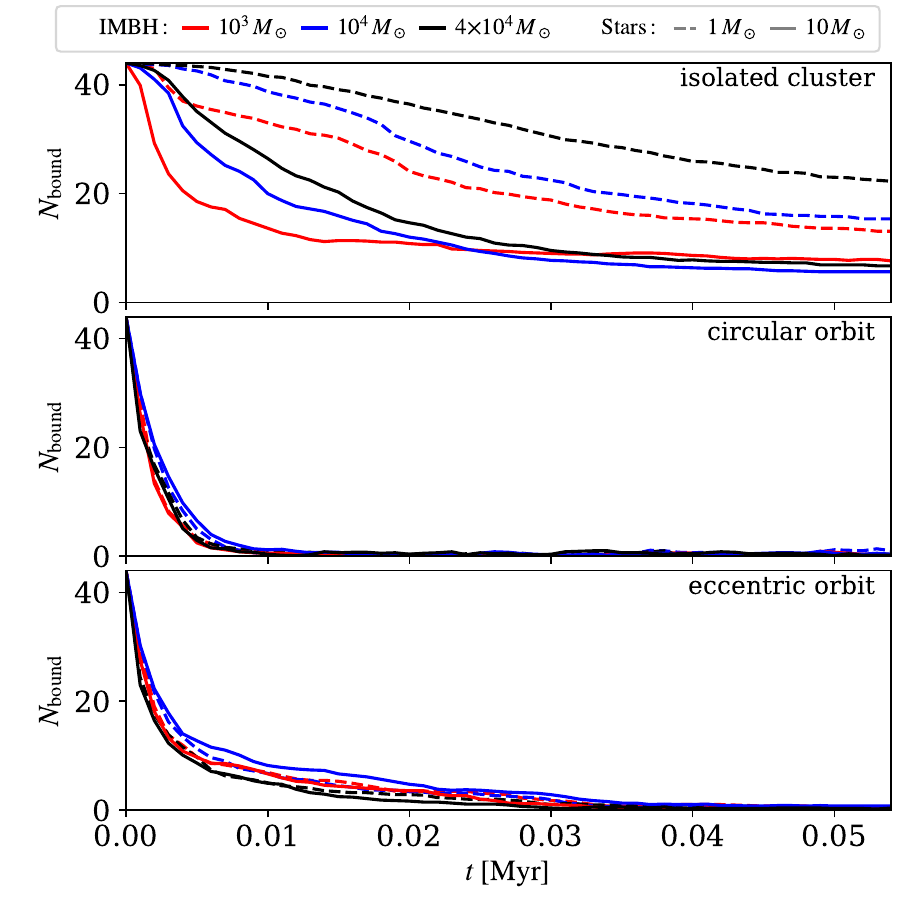}
    \caption{Time evolution of the number of stars bound to the IMBH (colour-coded based on its mass) in the SC models containing 44 stars initially and an IRS13-like IMBH. Each panel shows a different set of models --- either isolated or on a specified orbit. The line style (dashed or solid) corresponds to the stellar masses in the SC. The average from 10 realisations is shown.}
    \label{fig:mod44_N}
\end{figure}

\begin{figure}
    \includegraphics[width=\linewidth]{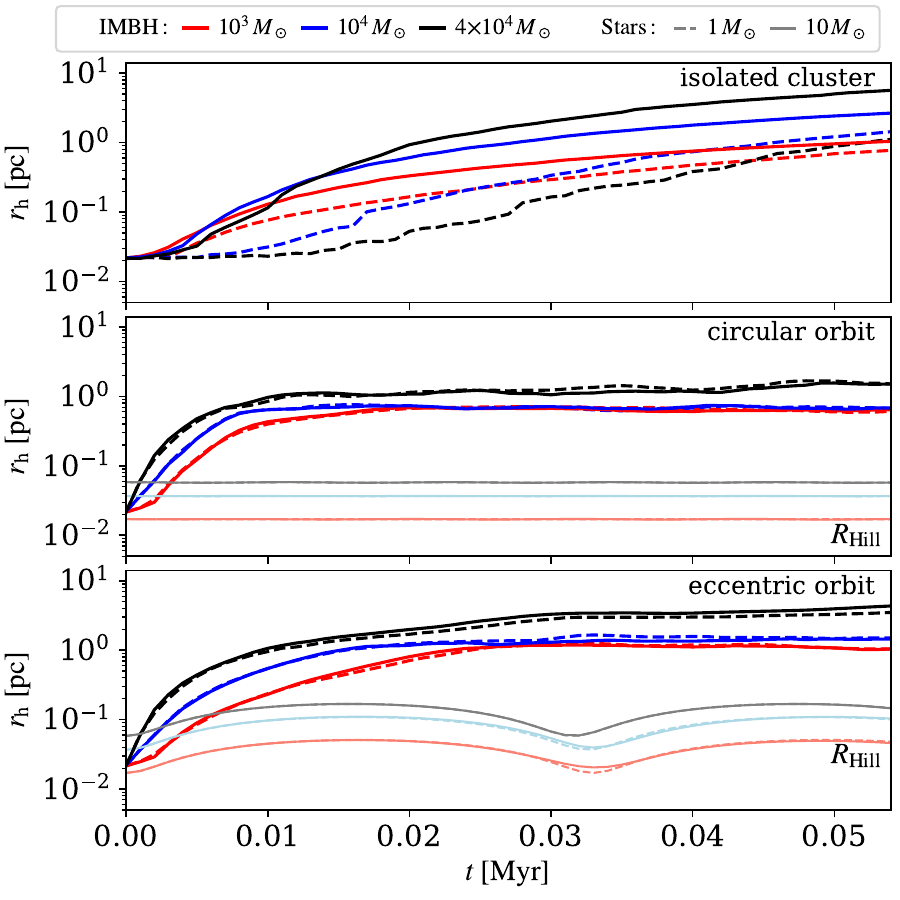}
    \caption{Similar to Fig.~\ref{fig:mod44_N} but for the time evolution of the half-mass radii. The Hill radius of each IMBH is drawn with a pale-coloured line of the corresponding model. The average from 10 realisations of each model is shown.}
    \label{fig:mod44_rh}
\end{figure}

\begin{figure}
    \includegraphics[width=\linewidth]{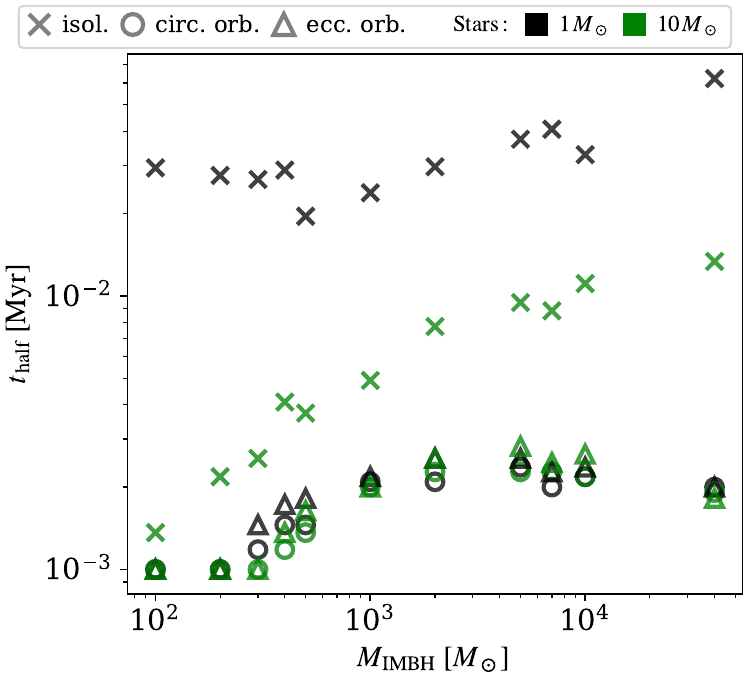}
    \caption{The time when the number of stars bound to the IMBH dropped by half (i.e.\ from 44 to 22). Different SCs are represented by their IMBH's mass on the horizontal axis. Different symbols indicate whether the SC is isolated, on a circular orbit, or eccentric orbit around the SMBH. The stellar masses are represented with colours.}
    \label{fig:thalf}
\end{figure}

\begin{figure}
    \includegraphics[width=\linewidth]{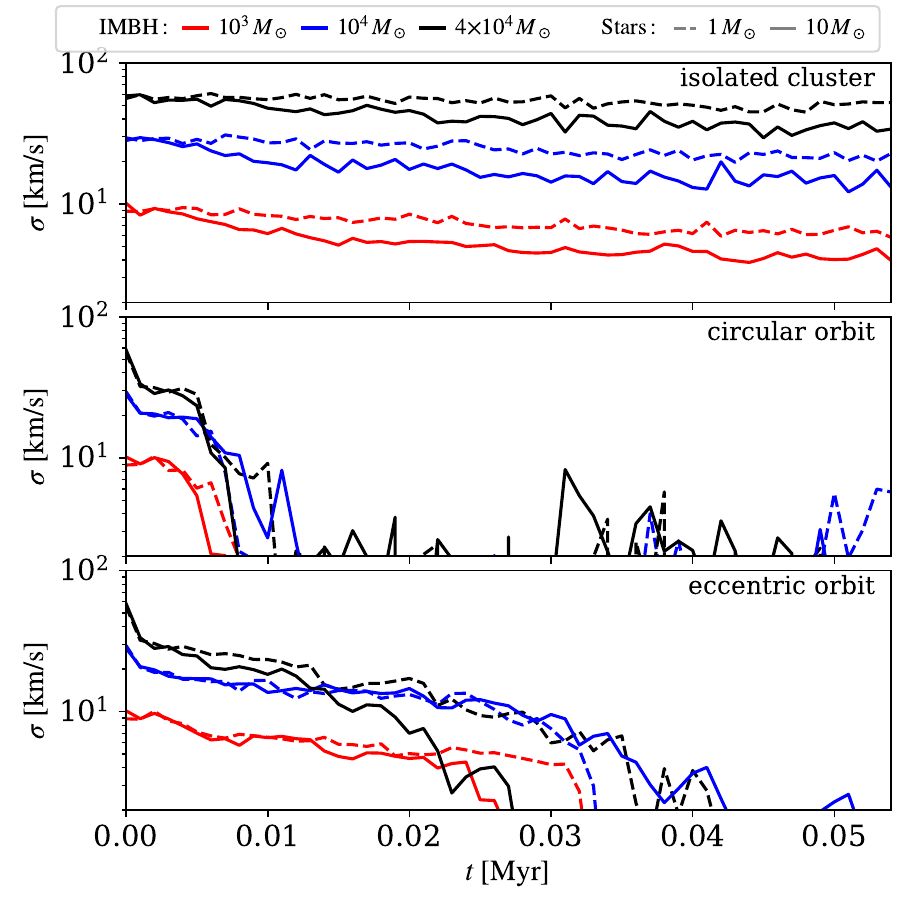}
    \caption{Similar to Fig.~\ref{fig:mod44_N} but for the time evolution of the velocity dispersion. The average from 10 realisations of each model is shown.}
    \label{fig:mod44_disp}
\end{figure}

\subsection{Results}

We find that all orbiting SCs expand and dissolve within a fraction of Myr \citep[i.e.\ comparable to their orbital period, which is consistent with][]{peissker_etal2024}. This result is similar regardless of the stellar masses or the mass of the IMBH. However, we notice a slightly slower rate of dissolution in the models on initially eccentric orbits (compare the same lines between the bottom two panels in Fig.~\ref{fig:mod44_N} and in Fig.~\ref{fig:mod44_rh}). Although the IRS13-size SC was initially bound, and in the case of the most massive IMBH also fully embedded within its Hill radius%
\footnote{The Hill radius at time $t$ is given by
$r_\mathrm{Hill}(t) = d_\mathrm{S{-}I}(t) \Mimbh^{1/3} (3\Mimbh + 3\Msgra)^{-1/3},$
where $d_\mathrm{S{-}I}$ is the separation of the SMBH and IMBH.}\!\!,
even such a massive point-mass object cannot hold the stars together \citep[some more extended massive component would perhaps be necessary to bind such a SC, see also][]{vonF_etal2022, jia_etal2023, singhal2024}. In fact, the opposite effect is seen --- the main reason for the stellar escape is the interaction between stars and the IMBH. Shortly after the start of the simulations (${\lesssim}10^{-4}\,\Myr$), we note the formation of a hard star--IMBH binary in all models, which then starts dynamically heating the SC and enhances the dissolution.

We further illustrate the rates of SCs' dissolution in Fig.~\ref{fig:thalf} where we plot the evolutionary times when the number of stars bound to the IMBH decreased by half. Regardless of the IMBH mass (here we show a wider range of $\Mimbh$), the trend remains --- i.e.\ the dissolution of isolated SCs containing $1\,\Msun$ stars is an order of magnitude slower than when the same SCs orbit the SMBH (compare the black crosses in the figure with the other black-coloured points).\footnote{We note that our models also show that for the isolated SCs, the more massive IMBHs can retain about a dozen bound stars for ${\sim}2$\,Myr, whereas the orbiting SCs dissolve completely and much sooner (see, e.g.\ Fig.~\ref{fig:mod44_N}).}

In the isolated case, this dynamical heating is more efficient in the SCs containing the $10\,\Msun$ stars where the central binary can become harder (see the top panel of Figs.~\ref{fig:mod44_N} \&~\ref{fig:mod44_rh}), and also for the lower-mass IMBHs where the required escape velocity is smaller (see the top panel of Fig.~\ref{fig:mod44_N} and the low-mass part in Fig.~\ref{fig:thalf}). Since the orbiting SCs are tidally limited, even stars that do not reach the escape velocity are unbound from the IMBH when they travel beyond the Hill radius. Consequently, the dissolution is mainly driven by the SMBH--IMBH interaction and the difference in stellar escapes is not as pronounced between the models with $1\,\Msun$ and $10\,\Msun$. We show further evidence of this in Fig.~\ref{fig:thalf} where (i) the SCs with an IMBH below a few hundred $\Msun$ are destroyed almost immediately, and (ii) regardless of stellar mass even an IMBH of thousands of solar masses, cannot hold the SC together for more than a fraction of Myr (see the circles and triangles forming a plateau above $10^3\,\Msun$).

As expected, at a distance of $0.4\,\pc$ from the SMBH, the long-term survival of such a small cluster is not plausible. Thus our orbital models lead us to conclude that rather than a self-gravitating, relaxed and bound system, the object we identify as IRS13 is more likely a random present-day clumping of stars, potentially deposited into the vicinity of \SgrA\ from farther away \citep[see also the remarks in][]{maillard_etal2004,fujii_etal2008}. Consequently, estimating the mass of its hypothetical central IMBH purely based on the stellar velocity dispersions of its members may lead to higher uncertainties, and can only give its upper mass limit.

To support this result further, we plot the 3D velocity dispersion of the bound stars in Fig.~\ref{fig:mod44_disp}. Let us emphasise that these modelled SCs suffered dissolution despite having five to ten times lower initial stellar velocity dispersion than the real IRS13 cluster.

\section{Infalling clusters}
\label{sec:infall}

In the following, we test several scenarios of what could plausibly happen with more populous SCs infalling into the central parsec from distances orders of magnitude larger.

\subsection{Models}

We performed a series of $N$-body simulations of SCs of different sizes and numbers of stars falling into the central regions of the Milky Way Galaxy (MW) from the initial distance of about $10{-}100\,\pc$ on various trajectories.
We followed their evolution numerically using \textsc{PeTar} \citep{petar}, with \textsc{GalPy} \citep{galpy} to set the external potential. In the computations, we combined the bulge, the disk and the halo (as implemented in \textsc{GalPy}'s \texttt{MWPotential2014}) with a Keplerian central SMBH (mass of $\Msgra$). We also used \textsc{GalPy} to include a~time-evolving Keplerian potential representing an IMBH on a circular orbit around the SMBH ($\Mimbh = 4\times 10^4\,\Msun$, $r = 0.4\,\pc$, orbiting in the same plane as the SC) such that the IMBH feels the gravitational force of the other external potentials and the SC. We note that we did not consider the scenario in which the IMBH would be in the SC initially because of the formation time constraints presented above (see also Fig.~\ref{fig:imbh_literature}).

All stars were treated as point masses, and primordial binary stars were not included (however, the dynamical formation of binaries was permitted in the simulations). To minimise the number of free parameters, we did not include star formation, stellar evolution, and the gas component; all are planned for our follow-up research.
Considering the stellar composition of IRS13, we restricted the maximum mass of the initial mass function by $10\,\Msun$; we kept the lower limit at $0.08\,\Msun$ and used the slopes from \citet{kroupa}. The stellar distribution for all SCs followed \cite{king_model} with $W_0=6$ \citep[numerically created with \textsc{McLuster};][]{mcluster}.
Table~\ref{tab:infall_models} lists the initial conditions selected for each SC run.

\begin{table}
	\centering
	\caption{Initial conditions of the SC models.}
	\begin{tabular}{lrrrrc}
		\hline
		Model name & $N_0$            & $\rg[0]$       & $v_{y,0}$        & $\rh[,0]$       & fall in \\
                   & {\small$(10^3)$} & {\small$(\pc)$} & {\small$(\kms)$} & {\small$(\pc)$} & ${<}0.5\,\pc$\\
		\hline\hline
        \texttt{20k-10-v5}      &  20 &  10 &  5.0 & 0.5 & \checkmark \\
        \texttt{20k-10-v10}     &  20 &  10 & 10.0 & 0.5 & \checkmark \\
        \texttt{20k-10-v20}     &  20 &  10 & 20.0 & 0.5 &  \\
        \texttt{50k-10-v5}      &  50 &  10 &  5.0 & 0.5 & \checkmark \\
        \texttt{50k-10-v10}     &  50 &  10 & 10.0 & 0.5 & \checkmark \\
        \texttt{50k-20-v10}     &  50 &  20 & 10.0 & 0.5 &  \\
        \texttt{100k-50-v5}     & 100 &  50 &  5.0 & 0.5 &  \\
        \texttt{100k-50-v5-w}   & 100 &  50 &  5.0 & 2.0 &  \\
        \texttt{100k-100-v1}    & 100 & 100 &  1.0 & 0.5 &  \\
        \texttt{100k-100-v1-w}  & 100 & 100 &  1.0 & 2.0 &  \\
        \texttt{100k-100-v5}    & 100 & 100 &  5.0 & 0.5 &  \\
        \texttt{100k-100-v5-w}  & 100 & 100 &  5.0 & 2.0 &  \\
        \texttt{100k-100-v10}   & 100 & 100 & 10.0 & 0.5 &  \\
        \texttt{100k-100-v10-w} & 100 & 100 & 10.0 & 2.0 &  \\
        \hline
	\end{tabular}
	\tablefoot{The columns show the number of stars, Galactocentric distance, velocity, and half-mass radius. The last column indicates whether a significant fraction of the SC members managed to fall into the central half-parsec within the estimated lifetime of IRS13 (see also Fig.~\ref{fig:sc_infall}). The Galactic plane is in the $xy$-plane. 
    Naming convention:
    [$N_0$]-[$\rg[0]/\pc$]-\texttt{v}[$v_0/\kms$]-[\texttt{w} (for wider SCs with $\rh[,0]=2\,\pc$)]}
	\label{tab:infall_models}
\end{table}

\subsection{Results}

\begin{figure*}
    \centering
    \includegraphics[width=.32\linewidth]{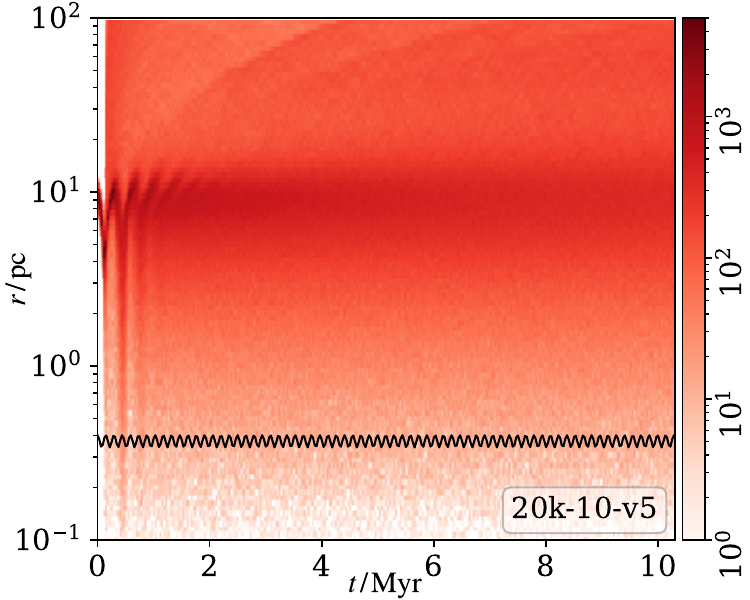}\hfill
    \includegraphics[width=.32\linewidth]{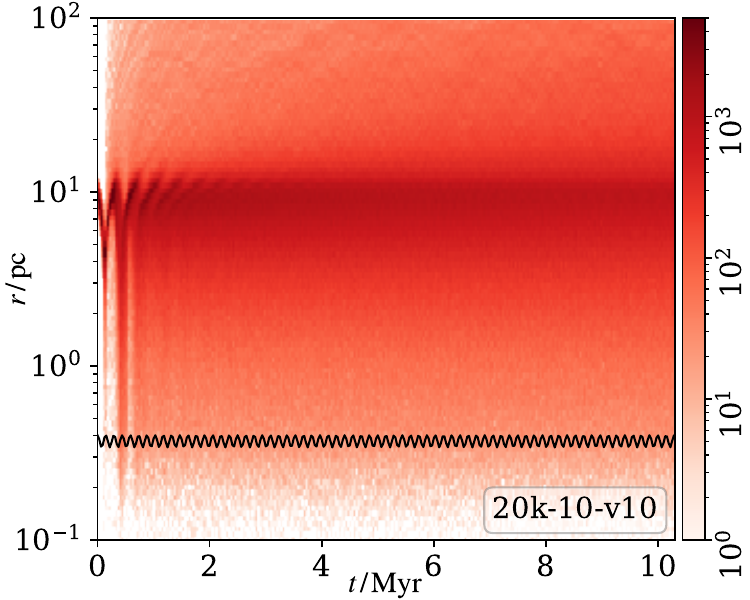}\hfill
    \includegraphics[width=.32\linewidth]{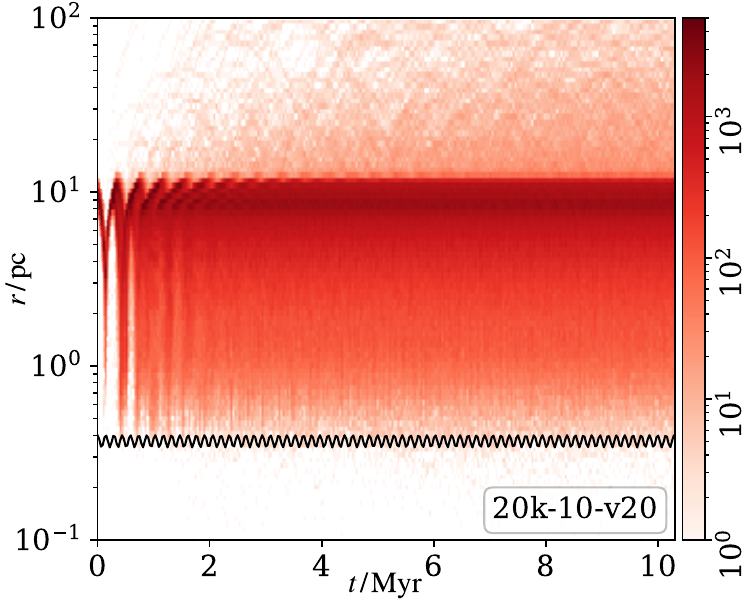}\\
    
    \includegraphics[width=.32\linewidth]{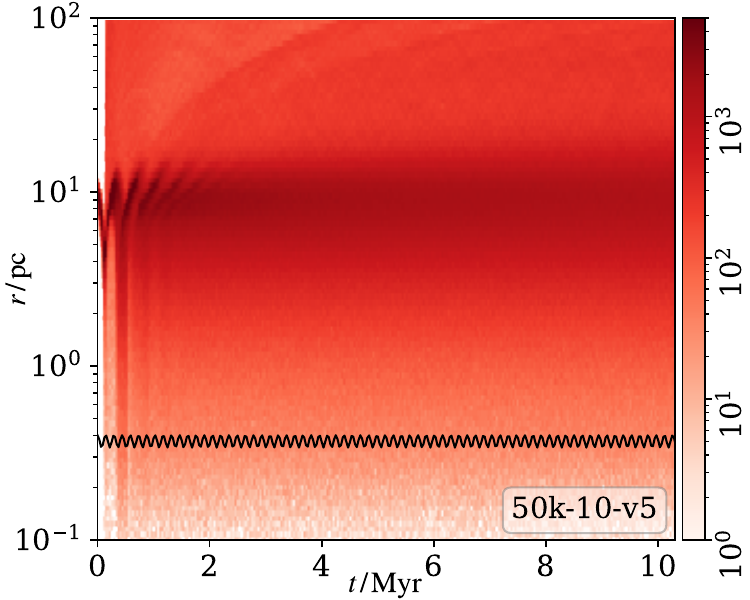}\hfill
    \includegraphics[width=.32\linewidth]{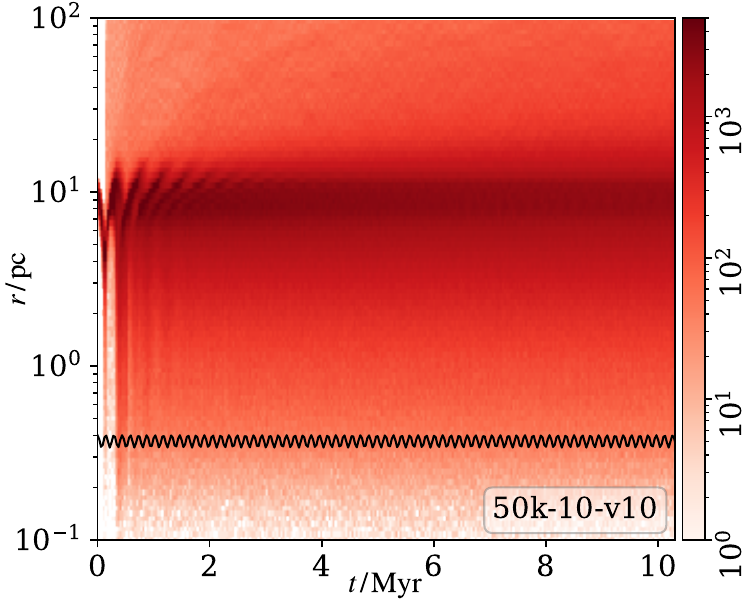}\hfill
    \includegraphics[width=.32\linewidth]{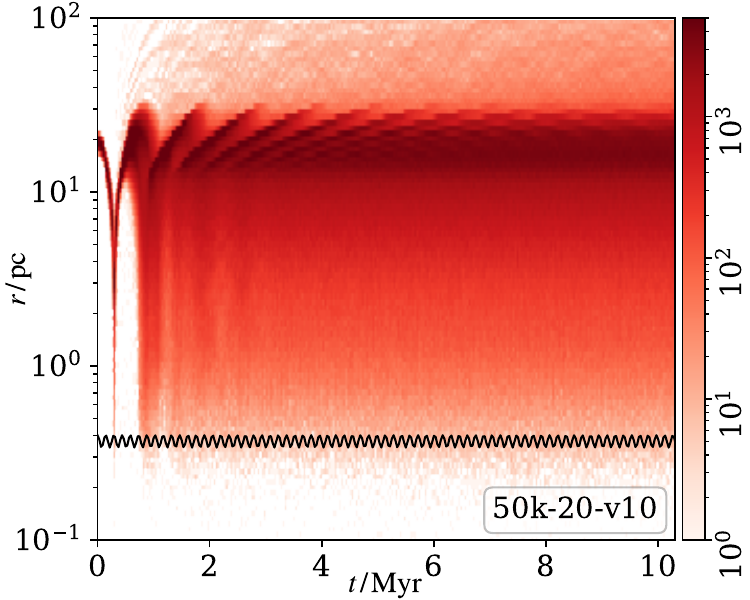}\\
    
    \includegraphics[width=.32\linewidth]{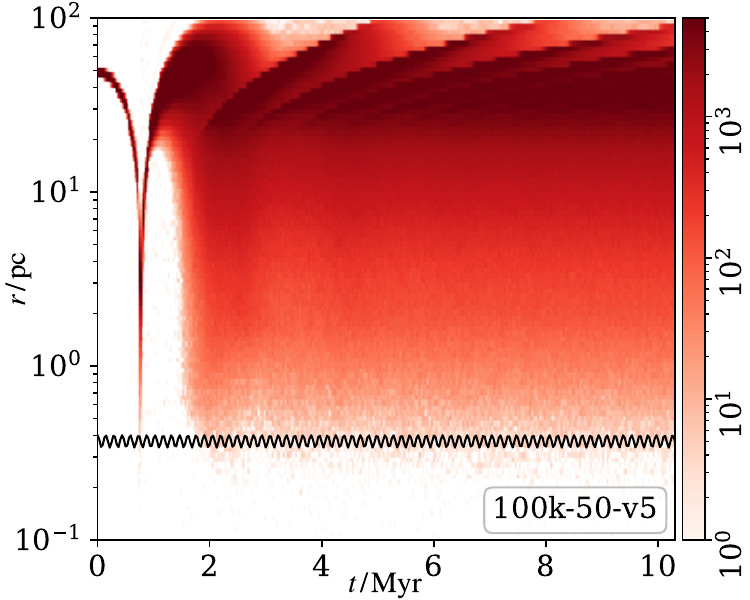}\hfill
    \includegraphics[width=.32\linewidth]{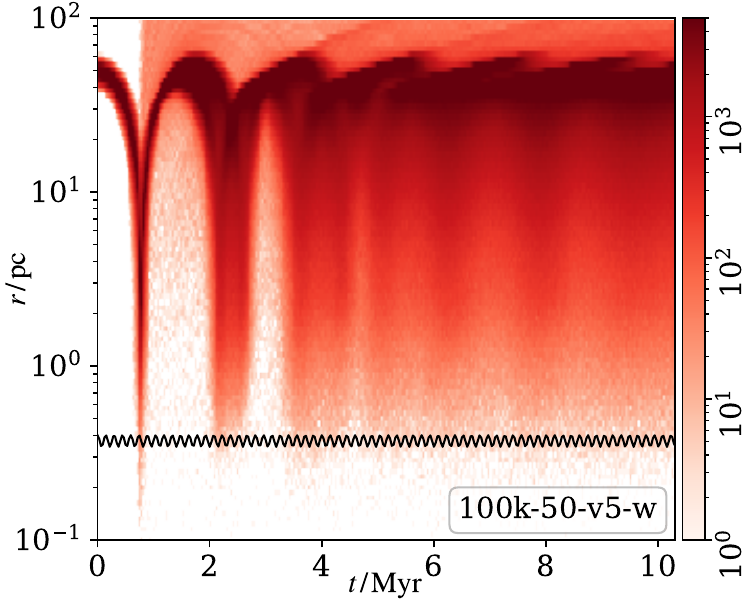}\hfill
    \includegraphics[width=.32\linewidth]{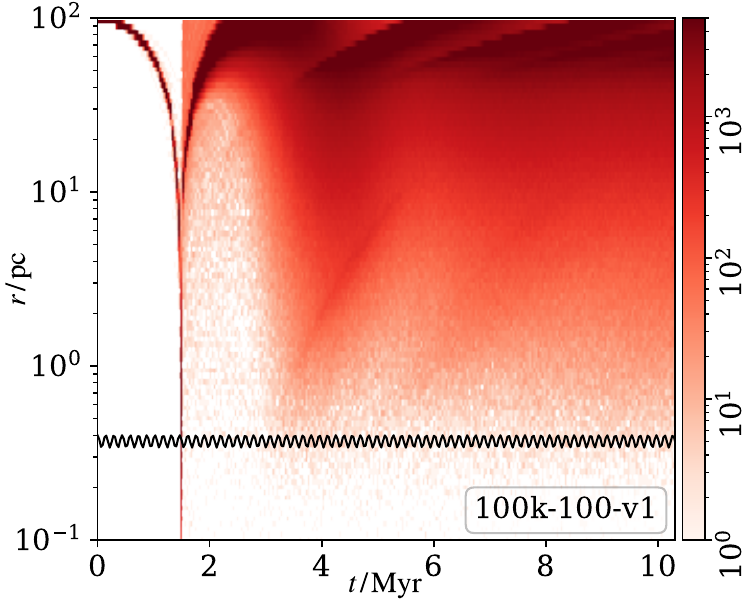}\\
    
    \includegraphics[width=.32\linewidth]{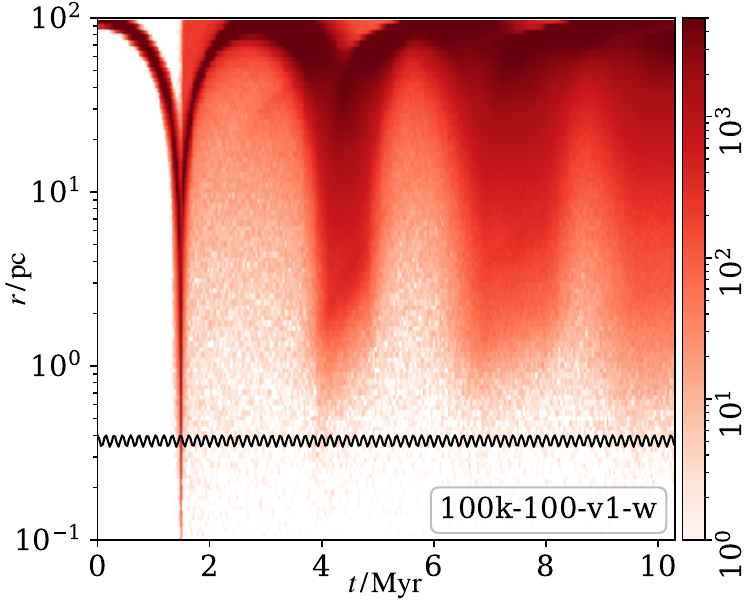}\hfill
    \includegraphics[width=.32\linewidth]{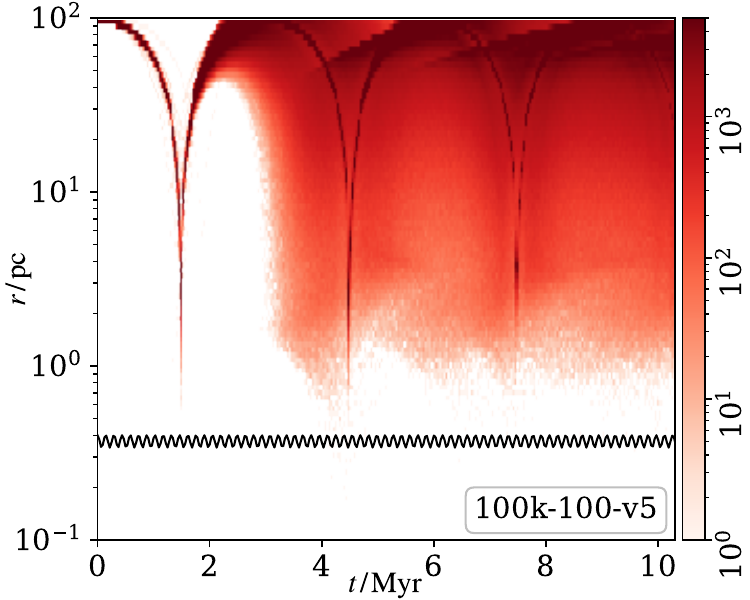}\hfill
    \includegraphics[width=.32\linewidth]{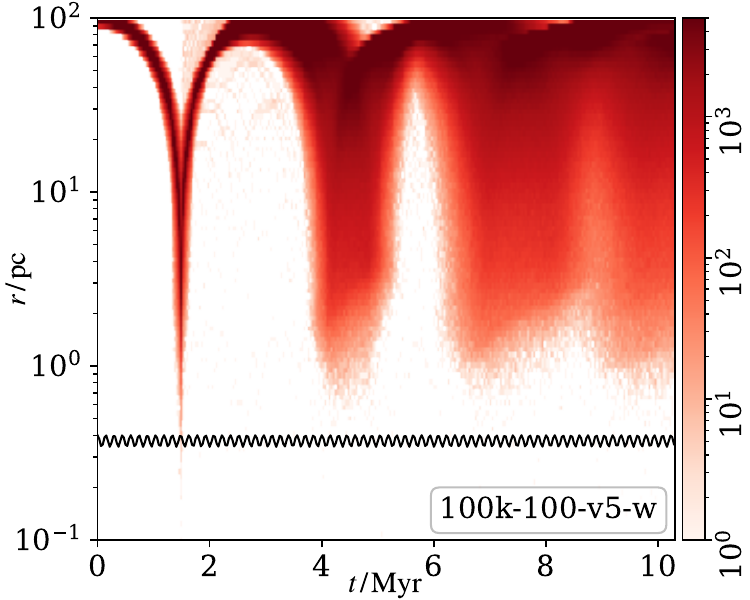}\\
    
    \includegraphics[width=.32\linewidth]{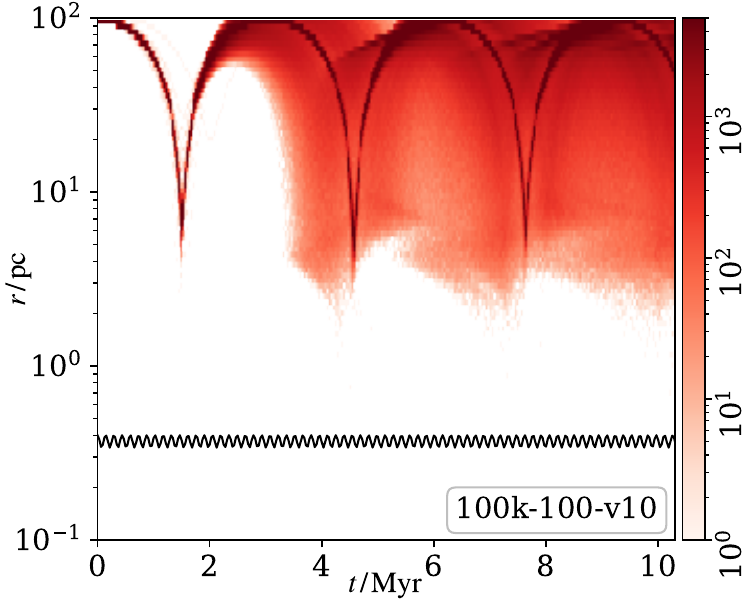}\hfill
    \includegraphics[width=.32\linewidth]{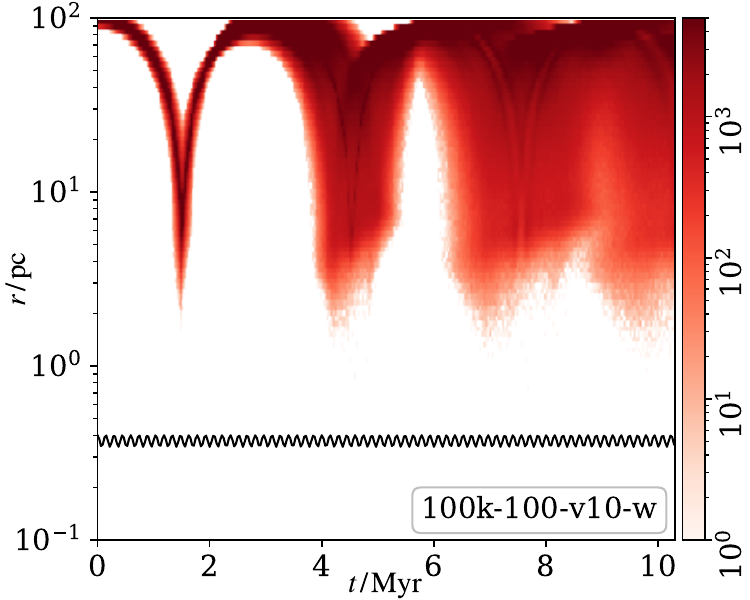}\hfill
    \begin{minipage}[b]{.32\linewidth}\caption{Time evolution of the infalling SCs and the emergence of temporary disk-like structures (see  Sect.~\ref{sec:infall} and Tab.~\ref{tab:infall_models} for the naming convention and model parameters). The vertical axis shows the Galactocentric distance, and the colour gradient corresponds to the number of stars in each pixel. The horizontal black dotted line is the Galactocentric distance of the IMBH. We emphasise that the size of the SC or its core is to be measured in the vertical direction. Downward pointing spikes are formed by stars that are brought close to the SMBH. At larger distances, the inclined overdensities constitute emerging ring-like features that originate from a gradually dissolving SC.}
    \label{fig:sc_infall}
    \end{minipage}
\end{figure*}

\begin{figure*}
    \centering
    \includegraphics[width=.495\linewidth]{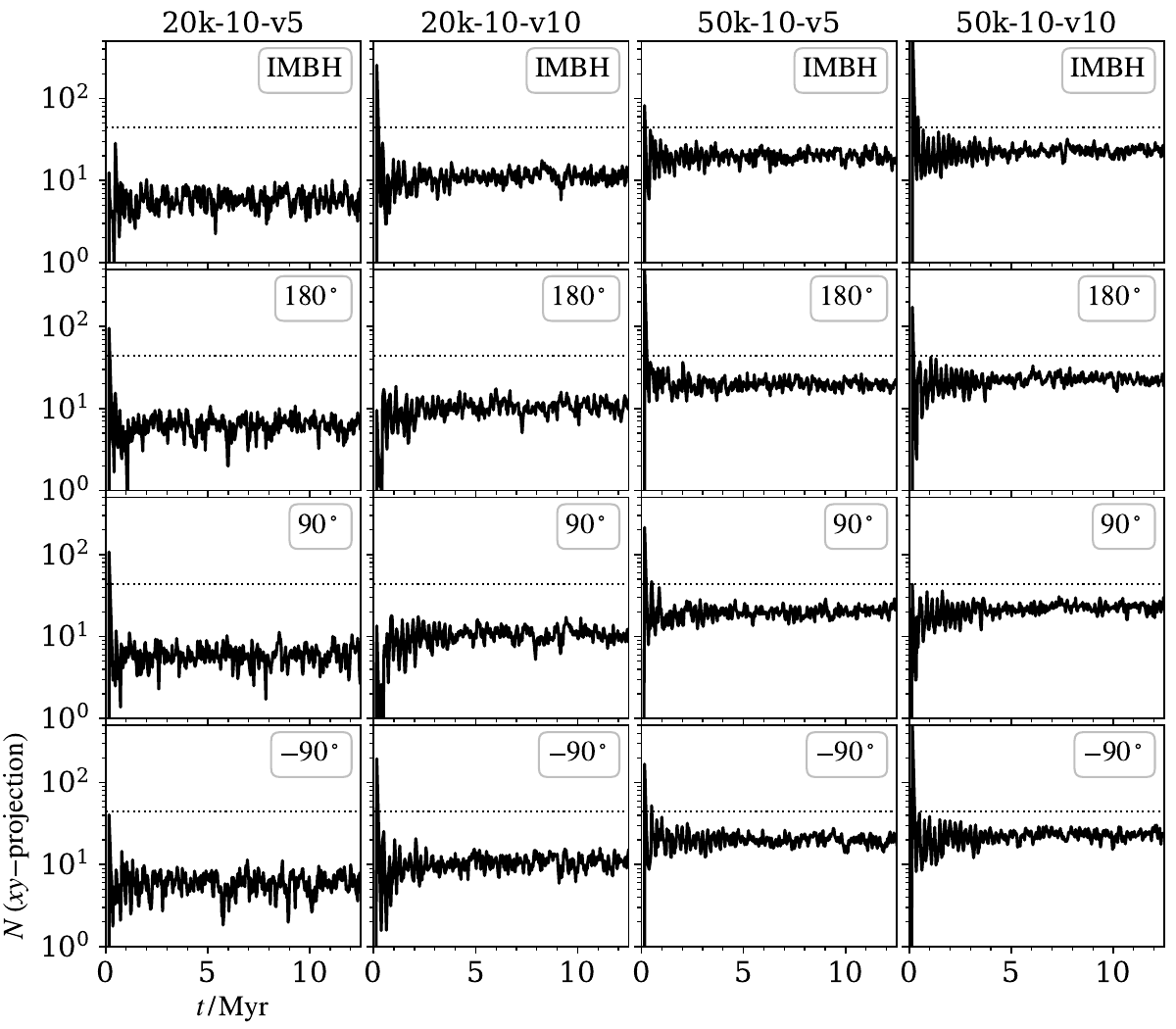}\hfill
    \includegraphics[width=.495\linewidth]{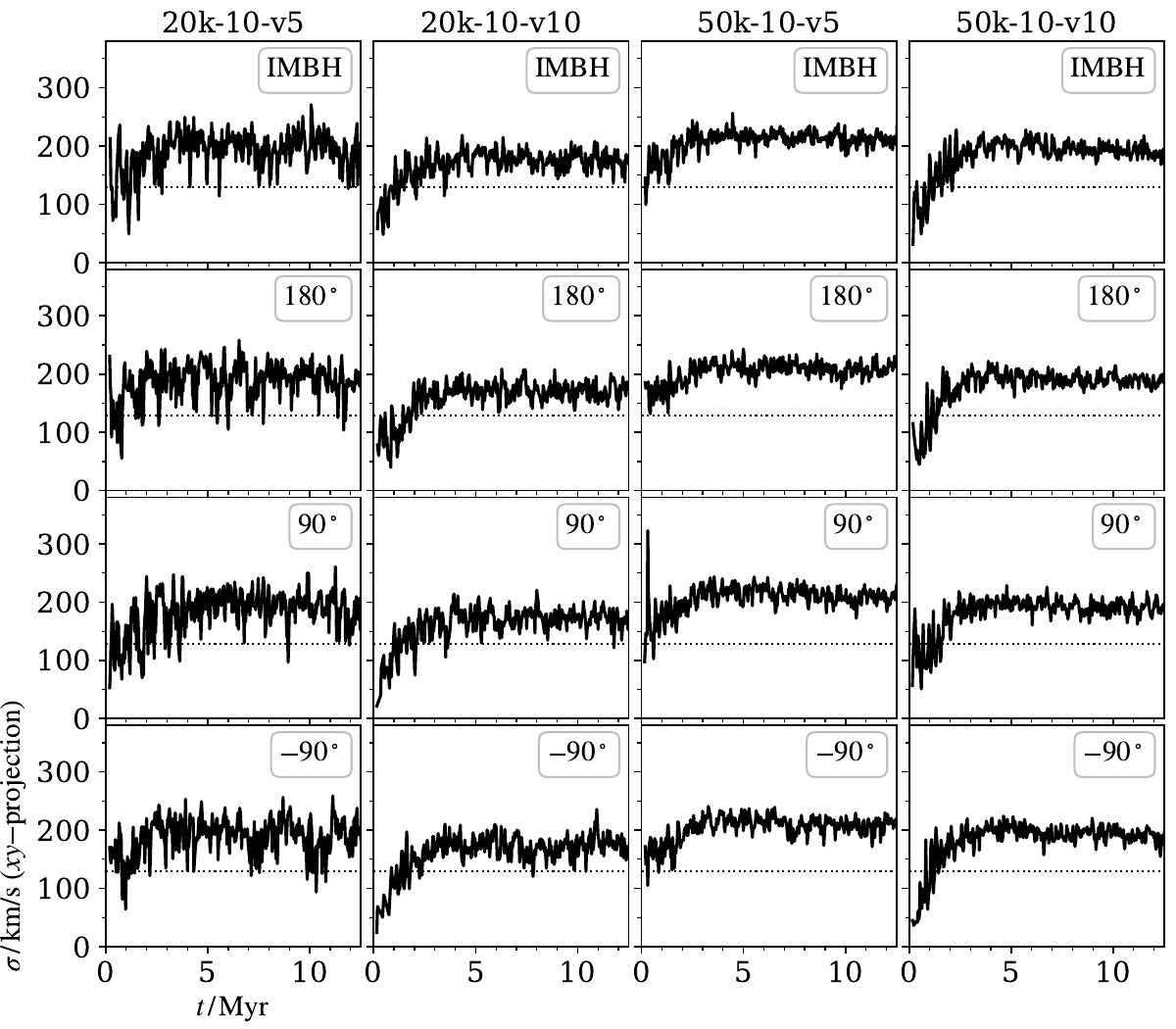}
    \caption{Time evolution of the number of stars in a specified region (left-hand plots) and their velocity dispersion (right-hand plots) as seen in $xy$-projection. For each model (columns), the values in four regions are shown --- around the IMBH (top row), on the opposite side of the SMBH than the IMBH (labelled $180\adeg$), and in two regions in the perpendicular direction to the SMBH--IMBH vector (labelled ${\pm}90\adeg$).
    The dashed lines show the known population and velocity dispersion of IRS13, respectively.}
    \label{fig:Nv_xy}

    \vspace{\floatsep}
    
    \includegraphics[width=.495\linewidth]{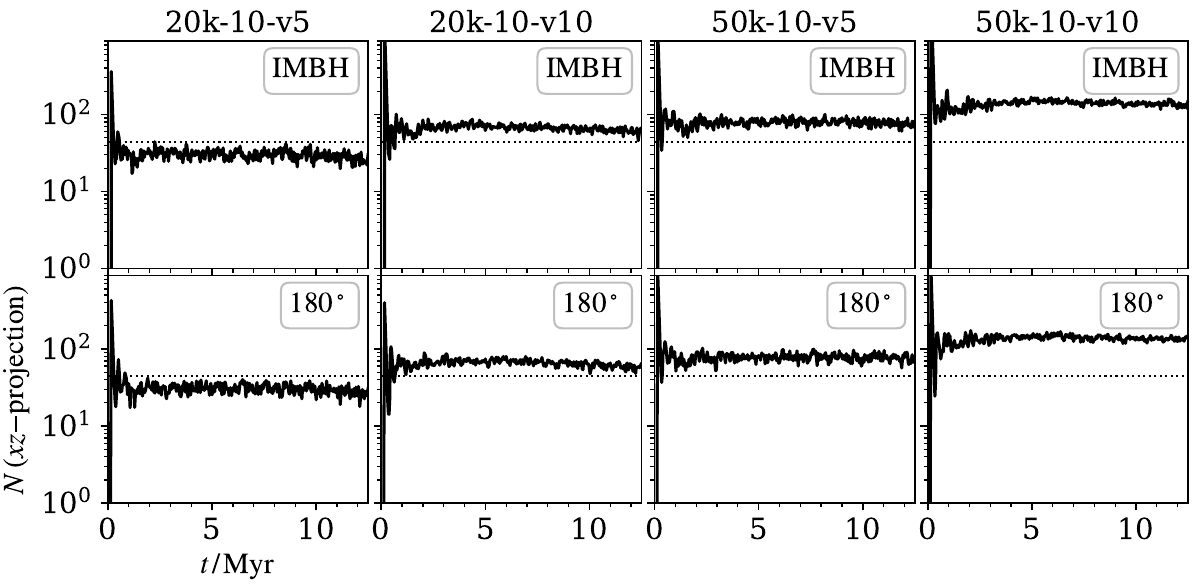}\hfill
    \includegraphics[width=.495\linewidth]{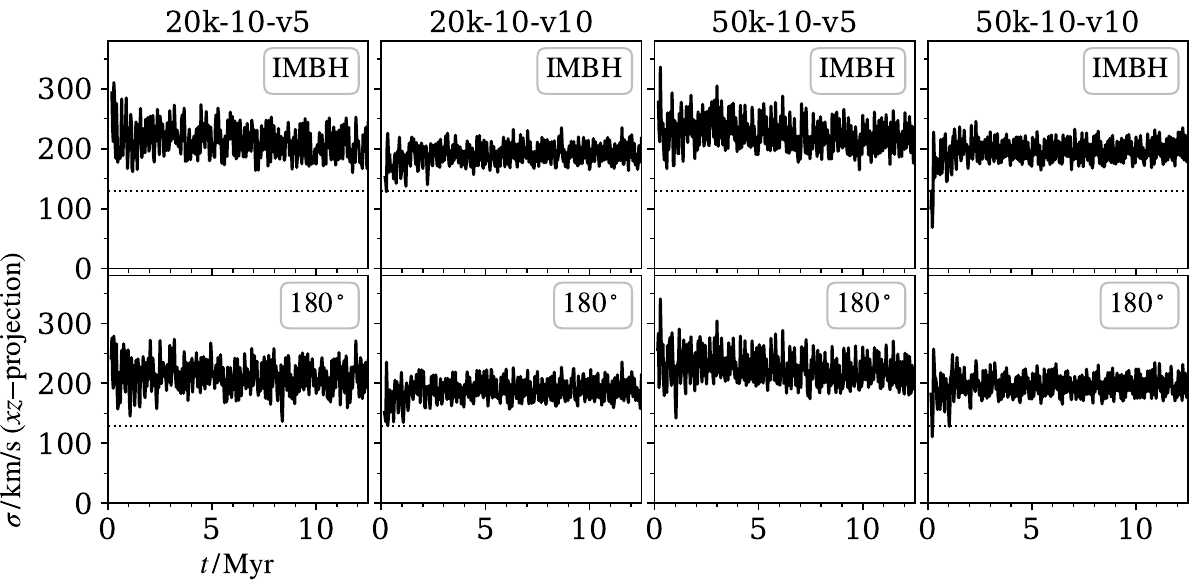}
    \caption{Same as Fig.~\ref{fig:Nv_xy} but for the $xz$-projection. Only two regions are shown --- around the IMBH (top row) and the region on the opposite side of the SMBH then the IMBH (labelled $180\adeg$).}
    \label{fig:Nv_xz}
\end{figure*}

\subsubsection{Tidal disruption of the SCs}

All infalling SCs show similar evolution (see the visualisations in Fig.~\ref{fig:sc_infall}). As they perform one or more orbits around the SMBH they are being stripped of stars and dissolved, forming disks and prominent spiral structures. The specific scenario and its time scale depend on several factors.

The first one is the trajectory (i.e.\ the farther away the SC orbits from the SMBH, the longer its core survives). Compare, for instance, the models \texttt{100k-100-v10} and \texttt{100k-100-v5} in Fig.~\ref{fig:sc_infall} (both are the same SCs, except the former has higher initial velocity and thus lower eccentricity); the SC's core is visible as the thinner arching line in each plot. The core of the model \texttt{100k-100-v5} becomes almost fully dissolved during its third visit to the SMBH whereas the core of \texttt{100k-100-v10} seems intact. We note that since the SC loses stars, which carry away angular momentum, the core does not follow a closed ellipse but a Rosetta orbit, forming several spiral-like structures depending on the number of passages through the pericentre.

The second factor is the compactness of the SC (i.e.\ its initial half-mass radius), however, the concrete scenario here also depends on the orbit. For the SCs with pericentres farther from the SMBH --- compare \texttt{100k-100-v5} and \texttt{100k-100-v10} with their wider (\texttt{-w}) counterparts in Fig.~\ref{fig:sc_infall} --- the cores of the initially wider models become even more tidally extended already during their second passage near the SMBH (see how the high-density line becomes very fuzzy; in the case of \texttt{100k-100-v10-w}, the split into two fainter cores is also visible). For the SCs that fall much closer to the SMBH the opposite trend is observed, i.e.\ the initially compact models suffer total disruption during the first passage near the SMBH whereas the wider SCs expand but do not dissolve completely --- compare \texttt{100k-50-v5} and \texttt{100k-100-v1} with their \texttt{-w} counterparts in Fig.~\ref{fig:sc_infall} (the plots of the wide SCs display several orbits but the initially compact SCs show smooth colour gradient).

The formation of the disk-like structures is also linked with the SC dissolution. If the SC dissolves completely, a stronger circular (or semi-circular) overdensity forms and slowly expands away from the SMBH. This can be seen very clearly in Fig.~\ref{fig:sc_infall} in the models \texttt{50k-20-v10}, \texttt{100k-50-v5} and \texttt{100k-100-v1} as the sharp, dark bows at higher radii. As the system relaxes, these expanding concentric structures mix with other stars and merge into a singular denser circle with a radius similar to the initial position of the SC. The time scale for this process is shorter when the number of stars in the system is smaller (compare, e.g.\ \texttt{20k-10-v20}, \texttt{50k-20-v10} and \texttt{100k-50-v5}). The strength of this final disk depends on the initial orbit and the dissolution rate --- compare, e.g.\ \texttt{20k-10-v5} (with the smallest density), \texttt{20k-10-v10} and \texttt{20k-10-v20} (with the highest density), or similarly the models \texttt{50k-10-v5}, \texttt{50k-10-v10} and \texttt{50k-20-v10}.

\subsubsection{Influence of the IMBH}

The SCs must have favourable initial conditions in order to deposit enough stars into the central regions of the MW. They either have to be sufficiently close (tens of $\pc$) or experience more radial orbits if they are farther away --- compare the stellar densities at the IMBH's orbit in Fig.~\ref{fig:sc_infall}.
This is consistent with \citet{kim_morris03} who studied the inspiral of SCs due to dynamical friction from ${<}30\,\pc$ and found that only the densest (${\gtrsim}10^6\,\Msun\,\pc^{-3}$) and most massive SCs may contribute to the stellar population in the central parsec. Similar results were also reported by \cite{fujii_etal2008} and \citet{petts_gualandris17}.
We mark the models that managed to fall inside the inner half-parsec around the SMBH in Table~\ref{tab:infall_models} and are further only focusing on them.

To mimic observations, we analyse our data in 2D projections --- either perpendicular to the Galactic plane (model coordinates $xy$, see Fig.~\ref{fig:Nv_xy}) or from within the Galactic plane (coordinates $xz$, see Fig.~\ref{fig:Nv_xz}). We evaluate the impact of the IMBH on the stellar motions of a tidally disrupted SC by focusing on four regions along the IMBH's orbit. Those are circular areas of radii $0.1\,\pc$, located (1) around the instantaneous position of the IMBH (i.e.\ $x{=}x_\bullet$, $y{=}y_\bullet$, $z{=}z_\bullet$); (2) on the opposite side of the SMBH than the IMBH (i.e.\ $x{=}{-}x_\bullet$, $y{=}{-}y_\bullet$, $z{=}{-}z_\bullet$); and (3\,\&\,4) in the perpendicular directions from the SMBH, but we show these only in the $xy$-projection (i.e.\ $x{=}{\pm}y_\bullet$, $y{=}{\mp}x_\bullet$). Hence, all regions are at the same distance from the SMBH but should feel different influences from the IMBH.

Due to the higher level of foreground and background contamination, the number of stars in each area is higher in the $xz$-projection than in the $xy$-projection (compare the left-hand panels of Figs.~\ref{fig:Nv_xy} \&~\ref{fig:Nv_xz}). In the \texttt{50k-10-v5} and \texttt{50k-10-v10} models, the number of stars is comparable to the population of IRS13; in the other two models, the values are lower, especially in the $xy$-projection.

We also show in the right-hand panels of Figs.~\ref{fig:Nv_xy} \&~\ref{fig:Nv_xz} that regardless of the position of the IMBH, stars in all regions have the same velocity dispersion in a given projection. Moreover, except for the $\sigma_{xy}$ value during the initial passage of the SC through the MW core, the velocity dispersion is always in the range $100{-}300\,\kms$, hence similar or higher than the measured value in IRS13. This reveals that at a distance of $0.4\,\pc$ from the SMBH, the stellar velocity dispersion of a tidally dissolved cluster is mainly driven by the SMBH, not an IMBH.

Our models also indicate a high mixing of stars in the central half-parsec. Therefore the stellar associations observed at one time snapshot usually do not survive together for more than a few thousand years.

\section{Conclusions}
\label{sec:concl}

We have modelled the stability of stellar clusters and associations in the gravitational field of the SMBH in the Milky Way Galaxy, taking into account the influence of the extreme tidal field in the neighbourhood of the central SMBH. We showed that an IRS13-like association is likely not a self-gravitating cluster, and even the presence of an IMBH of $4{\times}10^4\,\Msun$ cannot ensure its long-term stability and does not prevent its gradual dissolution in the time-frame of a few thousand years \citep[as is also seen in the observations, see][]{peissker_etal2023, peissker_etal2024}.

We have revisited arguments from the literature against the existence of such a massive IMBH existing so close to the \SgrA. Based on numerical modelling of the growth of IMBHs, we also emphasise that the origin of an IMBH of this mass is probably not the same as that of IRS13 --- see also \citet{takekawa_etal2017} who found two gaseous clouds in the central region of our Galaxy likely containing black holes that plunged into them.

We further showed that if IRS13 is a remnant of a tidally disrupted SC, the velocity dispersion of its members is determined mainly by the infall event and the SMBH. The IMBH's effect is only secondary or even negligible. Consequently, we question whether the estimates of a potential IMBH in IRS13 based on stellar kinematics are reliable.

Although we gained new insights into the dynamical interplay between SCs, IMBH and SMBH, we are aware that our current results are based only on simulations performed within the $N$-body approach with an external tidal field. Since the time scale for the dynamical stability of IRS13 is so short, this conclusion would likely remain unchanged even with the background gas, but the infall of SCs may play out differently. Therefore, we also aim to include the hydrodynamical treatment of gas and star formation to assess the potential limitations of the present approach (this is currently a work in progress which remains beyond the scope of the current work).

It is also of interest to note that arguments similar to those we put forward in this paper may apply to IRS1W dense stellar association which is located on the opposite side of \SgrA\ with respect to IRS13 and in a comparable distance from SMBH \citep{hosseini2022,hosseini2023}.
This confirms that the Galactic Centre is a promising location to explore tentative formation mechanisms of IMBHs and to set further constraints on their presence.

\begin{acknowledgements}
We thank Richard W\"unsch and Michal Zaja\v{c}ek for valuable discussions, Long Wang for help with \textsc{PeTar}, and Eugene Vasiliev for help with \textsc{Agama}.
VP has received funding from the European Union's Horizon Europe and the Central Bohemian Region under the Marie Skłodowska-Curie Actions -- COFUND, Grant agreement \href{https://doi.org/10.3030/101081195}{ID~101081195} (``MERIT'').
Views and opinions expressed are, however, those of the authors only and do not necessarily reflect those of the European Union or the Central Bohemian Region. Neither the European Union nor the Central Bohemian Region can be held responsible for them.
VP is also grateful for the access to
the computational resources provided by the e-INFRA CZ project (ID:90254), supported by the Ministry of Education, Youth and Sports of the Czech Republic;
and the computational cluster VIRGO at the Astronomical Institute of the Czech Academy of Sciences;
VK thanks the Czech Science Foundation grant GM24-10599M.
VP and VK also acknowledge the support from the project RVO:67985815 at the Czech Academy of Sciences.
FP gratefully acknowledges the Collaborative Research Center 1601 funded by the Deutsche Forschungsgemeinschaft (DFG, German Research Foundation) -- SFB 1601 [sub-project A3] -- 500700252.
The \texttt{Python} programming language with \texttt{NumPy} \citep{numpy} and \texttt{Matplotlib} \citep{matplotlib} were used in this project.
The \textsc{Rebound}'s \texttt{SimulationArchive} format was used to store fully reproducible simulation data \citep{reboundsa}.
This research has made use of NASA's Astrophysics Data System Bibliographic Services.
\end{acknowledgements}

\bibliographystyle{aa}
\bibliography{main}

\end{document}